\documentstyle[aps,prl,twocolumn,psfig,floats]{revtex}     
\begin{document}
\preprint{\vbox{\hbox{{\tt hep-ph/0312173}\\ December 2003}}}
\draft
\wideabs{
\title{Thermal Duality Confronts Entropy:  \\
A New Approach to String Thermodynamics?}
\author{Keith R. Dienes \,and\, Michael Lennek}
\address{Department of Physics, University of Arizona, Tucson, AZ  85721 USA}
\address{E-mail addresses:  ~{\tt dienes,mlennek@physics.arizona.edu}}
%  \date{December 16, 2003}
\maketitle
\begin{abstract}
One of the most intriguing features of string thermodynamics is
thermal duality, which relates the physics
at temperature $T$ to the physics at inverse temperature $1/T$.
Unfortunately, the traditional definitions of thermodynamic quantities
such as entropy and specific heat are not invariant under
thermal duality transformations.  In this paper, we discuss
several novel approaches towards dealing with this issue.  
One approach yields a ``bootstrap'' method 
of extracting possible exact, closed-form 
solutions for finite-temperature
effective potentials in string theory.  Another approach 
involves extending the usual definition of entropy by 
introducing additional terms which are suppressed by powers of the 
string scale.  At high temperatures, however, these string corrections
become significant and lead to a variety of surprising new phenomena 
as the temperature is increased.  
\end{abstract}
%  \pacs{11.25.-w,05.70.-a,04.70.Dy,05.90.+m}
\bigskip
\bigskip
          }

%========================================================================
%          KEYSROKE-SAVING MACROS, nothing complicated 
%========================================================================
\newcommand{\newc}{\newcommand}
\newc{\gsim}{\lower.7ex\hbox{$\;\stackrel{\textstyle>}{\sim}\;$}}
\newc{\lsim}{\lower.7ex\hbox{$\;\stackrel{\textstyle<}{\sim}\;$}}

\def\beq{\begin{equation}}
\def\eeq{\end{equation}}
\def\beqn{\begin{eqnarray}}
\def\eeqn{\end{eqnarray}}
\def\calM{{\cal M}}
\def\calV{{\cal V}}
\def\calF{{\cal F}}
\def\half{{\textstyle{1\over 2}}}
\def\ie{{\it i.e.}\/}
\def\eg{{\it e.g.}\/}
\def\etc{{\it etc}.\/}

%     The following macros are to create the "blackboard bold"
%     characters for "R" (set of real numbers),
%     "C" (set of complex numbers), and "Q" (set of rational numbers).

\def\inbar{\,\vrule height1.5ex width.4pt depth0pt}
\def\IR{\relax{\rm I\kern-.18em R}}
 \font\cmss=cmss10 \font\cmsss=cmss10 at 7pt
\def\IQ{\relax{\rm I\kern-.18em Q}}
\def\IZ{\relax\ifmmode\mathchoice
 {\hbox{\cmss Z\kern-.4em Z}}{\hbox{\cmss Z\kern-.4em Z}}
 {\lower.9pt\hbox{\cmsss Z\kern-.4em Z}}
 {\lower1.2pt\hbox{\cmsss Z\kern-.4em Z}}\else{\cmss Z\kern-.4em Z}\fi}
%========================================================================

\input epsf

%========================================================================
%========================================================================
%               MAIN TEXT BEGINS HERE
%========================================================================

%========================================================================
\section{Introduction}

Some of the most intriguing features of string theory have been
the existence of numerous dualities which connect physics in
what would otherwise appear to be vastly dissimilar regimes.
Such dualities include strong/weak coupling duality (S-duality)
as well as large/small compactification radius duality (T-duality),
and together these form the bedrock upon which much of our
understanding of the full, non-perturbative moduli space
of string theory is based.

There is, however, an additional duality which has received far
less scrutiny:  this is {\it thermal duality}\/, which relates
string theory at temperature $T$ with string theory at the inverse temperature
$T_c^2/T$ where $T_c$ is a critical (or self-dual) temperature related
to the string scale.  Thermal duality follows naturally from T-duality
and Lorentz invariance, and thus has roots which are as deep
as the dualities that occur at zero temperature.
Given the importance of dualities of all sorts in
extending our understanding of the unique features of
non-perturbative string theory,
we are led to ask what new insights can be gleaned from
a study of thermal duality.

The observation that underpins our approach
is a simple one:   classical
thermodynamics, as traditionally formulated, is not invariant
under thermal duality.
While certain thermodynamic quantities such as the free energy
and the internal energy of an ideal closed string gas exhibit
invariances (or covariances) under thermal duality transformations,
other quantities such as entropy and specific heat do not.

It is, of course, entirely possible that thermal duality
should be viewed only as an ``accidental'' symmetry
of the string effective potential in cases where it arises;  we thus
would have no problem with the loss of this symmetry
when calculating certain thermodynamic quantities.
However, given the close association between thermal duality and the
other dualities of string theory, it 
seems more natural to consider thermal duality
as a fundamental property of a consistent string theory,
and demand that this symmetry hold for {\it all}\/ physically
relevant thermodynamic quantities.

As we shall see, imposing this extra requirement launches 
a variety of new approaches towards thinking about traditional
string thermodynamics.  In each approach, however, our goal is the
same:  to reconcile the apparent conflict between thermal duality
and the rules of standard thermodynamics.

\section{~Thermal duality confronts entropy:  ~The problem}

Let us begin by reviewing the source of the problem.
Classical thermodynamics provides us with a number
of physically relevant temperature-dependent observables:  
these include the free energy (or effective potential) $F$, 
the internal energy $U$, the
entropy $S$, and the fixed-volume specific heat $c_V$, all
of which 
descend from a single 
thermal partition function
or vacuum amplitude $\calV\equiv -\ln Z$ through relations of the
form
\beqn
          F = T \calV,~~~~~~~
         && U = - T^2 {d\over dT} \calV,~~~\nonumber\\
         S = -{d\over dT} F,~~~~~~~ && c_V = {d\over dT} U~.
\label{usualrelations}
\eeqn
Our task is then to calculate $\calV(T)$.
For example, in closed string theories,
the one-loop contribution to $\calV$ is given by a modular integral
of the form~\cite{Pol86,McClainRoth,OBrienTan,Polbook}
\beq
    \calV(T) ~\equiv~ -\half \,{\cal M}^{D-1}\, \int_{\cal F} {d^2 \tau\over ({\rm Im} \,\tau)^2}
             \,Z_{\rm string}(\tau,T)
\label{Vdef}
\eeq
where ${\cal M}\equiv M_{\rm string}/2\pi$ is the reduced string scale;
$D$ is the spacetime dimension;
$\tau$ is the complex modular parameter describing the shape of the
one-loop toroidal worldsheet;
${\cal F}\equiv \lbrace \tau:  |{\rm Re}\,\tau|\leq \half,
 {\rm Im}\,\tau>0, |\tau|\geq 1\rbrace$ is the fundamental domain
of the modular group;
and $Z_{\rm string}(\tau,T)$ is the appropriate thermal string
partition function.

However, one important feature that emerges in such
string calculations is thermal duality (for early papers, see 
  Refs.~\cite{OBrienTan,AlvOsoNPB304,AtickWitten,AlvOsoPRD40,OsoIJMP}).
It is easy to understand how this symmetry arises.
In string theory (just as in ordinary quantum field theory),
finite-temperature effects can be
incorporated~\cite{Pol86}
by compactifying an extra (Euclidean) 
time dimension on a circle (or orbifold~\cite{shyam1})
of radius $R_T = (2\pi T)^{-1}$.
The Matsubara modes are nothing but the Kaluza-Klein states corresponding
to this compactification.
However, Lorentz invariance guarantees that the properties of this
extra time dimension should be the same as those of the original space
dimensions, and T-duality tells us that closed string theory
on a compactified space dimension of radius $R$ is indistinguishable
from that on a space of radius $R_c^2/R$ where $R_c$ is a critical,
self-dual radius~\cite{SakaiSenda,Nairetal,Sathiapalan}.
Together, these two symmetries thus imply a thermal
duality symmetry (or thermal {\it self}\/-duality symmetry) 
under which $Z_{\rm string}$ 
(and therefore $\calV$) is invariant under
the thermal duality transformation $T\to T_c^2/T$:
\beq
      Z_{\rm string}(\tau,T_c^2/T) ~=~ 
      Z_{\rm string}(\tau,T)~.
\label{dualityrelation}
\eeq
In other words, thermal Matsubara modes 
are accompanied in string theory by thermal winding modes;
exchanges between the two sets of states
results in a duality symmetry which transcends
the behavior of point-particle quantum field theories.
Note that this symmetry exists to all orders 
in perturbation theory~\cite{AlvOsoPRD40}.

Given the emergence of thermal duality for closed strings,
the fundamental issue that shall concern us is
the failure of the rules in Eq.~(\ref{usualrelations}) to
respect this symmetry.   It is, of course, immediately evident
that $\calV$, $F$, and $U$ all continue to exhibit simple
thermal duality transformation properties 
as a result of the thermal duality invariance of $Z_{\rm string}$.
More specifically, we shall refer to 
a general thermodynamic quantity $f(T)$ as 
transforming covariantly 
with weight $k$ and sign $\gamma=\pm 1$
if $f(T_c^2/T) = \gamma (T_c/T)^k f(T)$ for all $T$.
We thus find that $\calV$, $F$, and $U$ each transform
covariantly with 
weights and signs $(k,\gamma)= (0,1)$, $(2,1)$, 
and $(2,-1)$ respectively. 
However, it is immediately apparent that $S$ and $c_V$ 
fail to transform covariantly (\ie, fail to close back
into themselves) under thermal duality.  Specifically, we find
\beqn
      S(T_c^2/T) &=& -S(T) - 2 F(T)/T~,\nonumber\\
      c_V(T_c^2/T) &=&  c_V(T) - 2U(T)/T~.
\label{failure}
\eeqn

Strictly speaking, this failure to transform covariantly does not
imply an inconsistency in either the rules of thermodynamics
or the existence of thermal duality.  Indeed, regardless of their form,
the transformation rules in Eq.~(\ref{failure}) 
provide a self-consistent mapping between quantities 
as measured by an observer using the temperature $T$
and a dual observer using the temperature $T_c^2/T$.

Nevertheless, it is very unnatural to find
a fundamental quantity such as entropy transforming non-covariantly
under such a duality transformation.   
If thermal duality is indeed a fundamental symmetry
of string theory, this suggests that entropy and specific
heat are improperly defined from a string-theoretic
standpoint.  At best, they are not the proper ``eigenquantities''
which should correspond to physical observables.
As an analogy, let us consider a system with gauge invariance.
Like thermal duality symmetries, gauge symmetries are really 
redundancies in description:
a system may be described in one gauge or another, just as 
a system may be described with certain
thermal modes labelled as ``momentum'' modes and others
labelled as ``winding'' modes, or vice versa.
However, in the case of gauge invariance, we know that 
physical quantities should be gauge-invariant,
transforming invariantly or covariantly under the gauge transformation.
By analogy, it therefore seems strange to have fundamental thermodynamic
quantities such as entropy which transform non-covariantly 
under duality transformations.
We shall find further support for this point of view below.

In this paper, we shall
therefore make the conjecture that 
the string-thermodynamic entropy should indeed  
be a covariant quantity, transforming covariantly under thermal
duality transformations. 
Our goal will then be to determine 
how we might reconcile this with the apparent results in Eq.~(\ref{failure}),
and what this might tell us about 
the thermal properties of string theory.

\section{~~Approach~\#1:\\   A thermal duality ``bootstrap''}

Our first approach towards addressing this issue is to investigate
whether there might nevertheless exist special
solutions for $\calV(T)$ such that duality covariance will be 
preserved for {\it all}\/ thermodynamic quantities, including
$S$ and $c_V$.  If so, then we can use the requirement of 
thermal duality covariance for $S$ and $c_V$ in order to
``bootstrap'' our way to special closed-form solutions for $\calV(T)$.
We would thus be exploiting thermal duality
in order to constrain the vacuum amplitude 
$\calV(T)$ in a manner that
transcends a direct order-by-order perturbative calculation.

It is straightforward to implement this bootstrap.
The reason that $S$ fails to be covariant, even when 
$F$ is covariant, is that the temperature derivative breaks
the duality covariance.
In general, if $f$ is a duality covariant function of weight $k$ and sign $\gamma$,
then
\beq
         \left\lbrack {df\over dT}\right\rbrack (T_c^2/T)
   =  - \gamma \left({ T_c\over T}\right)^{k-2} \left(
                     {df\over dT} - {k f\over T}\right) ~.
\label{derivbad}
\eeq
It is the final term on the right side of Eq.~(\ref{derivbad}) which 
generally prevents $df/dT$ from transforming as a covariant function when $k\not=0$.  
However, there is one special case when this does not pose a problem:
if 
\beq
           {df\over dT} - {k f\over T}~=~ -\delta  \left( {T_c\over T}\right)^\ell {df\over dT}~,
\label{constraint}
\eeq
for some sign $\delta=\pm 1$ and exponent $\ell$,
then $df/dT$ will be covariant, with 
weight $k+\ell-2$ and sign $\gamma \delta$. 
When can this happen?
Solving Eq.~(\ref{constraint}), we find that $f$ must have the general form
$f(T) \sim   (T^\ell + \delta   T_c^\ell )^{k/\ell}$ 
where $ \delta^{ k/\ell} = \gamma$.

Given this result, we can immediately determine which functional forms
$\calV(T)$ give rise to duality covariant entropy and specific heat. 
Since the free energy $F(T)$ has $(k,\gamma)=(2,1)$,
we learn that the entropy $S(T)$ will be duality covariant only
if $F(T)$ has the closed form
\beq
     F^{(\ell)}(T) ~\sim~ -{ (T^\ell + \delta T_c^\ell)^{2/\ell}\over T_c }~
\label{Fsoln}
\eeq
where $\delta^{2/\ell}=1$.  We shall henceforth choose $\delta=1$
since we do not expect $F(T)$ to vanish at $T=T_c$.
This implies that our remaining thermodynamic quantities must take
the closed forms 
\beqn
          \calV^{(\ell)}(T) &\sim& - (T^\ell + T_c^\ell)^{2/\ell}/ T T_c \nonumber\\
          U^{(\ell)}(T) &\sim& \phantom{-} (T^\ell + T_c^\ell)^{2/\ell-1}
               (T^\ell - T_c^\ell)/ T_c \nonumber\\
          S^{(\ell)}(T) &\sim&  2 \,T^{\ell -1} (T^\ell + T_c^\ell)^{2/\ell -1}/ T_c\nonumber\\
        c^{(\ell)}_V(T) &\sim&  2 \,T^{\ell-1} (T^\ell + T_c^\ell)^{2/\ell -2}\times\nonumber\\
         &&~~~~~~~~~~~\left\lbrack   T^\ell + (\ell -1) T_c^\ell \right\rbrack/ T_c~.
\label{lsoln}
\eeqn
As expected, all of these quantities are duality covariant except for 
$c_V$.  However, it is easy to verify that $c_V$ is covariant if $\ell=1$ or
$\ell=2$.
Thus, from amongst all possible duality covariant functions $\calV(T)$,
we have found that only the special closed-form solution listed above
enables all subsequent thermodynamic quantities to be 
duality covariant as well.

Of course, the real issue is to determine whether these functional
forms actually correspond to the results of explicit one-loop modular
integrations of the sort that
can emerge from actual finite-temperature duality covariant
string ground states.

Let us first concentrate on the $T\to 0$ and $T\to \infty$ limits.
In these limits, it is well known~\cite{AtickWitten,OsoIJMP,Polbook}
that we must have 
$F(T)\to \Lambda$ and $F(T)\to \Lambda T^2/T_c^2$ respectively,
where $\Lambda$ is the corresponding (zero-temperature) 
one-loop cosmological constant.  It is immediately apparent
that our closed-form solutions in Eq.~(\ref{Fsoln}) have these properties 
for all $\ell$, which in turn enables us to identify $\Lambda$ as 
the unknown normalization constant in Eqs.~(\ref{Fsoln}) and (\ref{lsoln}).

Another test is to determine whether our closed-form solutions
have the correct field-theoretic limit.  
Let us first recall that in $D$ spacetime dimensions, quantum field theory 
predicts $F(T)\sim T^D$ as $T\to \infty$.
This differs markedly from the expected 
string-theoretic asymptotic behavior $F(T)\sim T^2$ 
as $T\to \infty$;  this reduced exponent indicates 
that string theory has a significantly reduced number of degrees of freedom
at high temperatures compared with field theory~\cite{AtickWitten}.
However, since the high-temperature limit of field theory corresponds to 
the {\it low}\/-temperature $T\ll T_c$ limit
of string theory, we expect that we should 
observe $T^D$ scaling for $T\ll T_c$.

Our closed-form solutions have this property as well.  
Keeping the first subleading term in Eq.~(\ref{Fsoln}), we find
      $F^{(\ell)}(T) \sim  \Lambda + {2\Lambda\over \ell}
               ( {T/T_c} )^\ell$ 
for $T\ll T_c$.
This enables us to identify $\ell=D$, which completely fixes
all of the free parameters in our functional forms.  Specifically,
we find 
\beq
           F(T) ~=~ \Lambda \left[  1+ \left( {T\over T_c}\right)^D\right]^{2/D}~,
\label{Ffullform}
\eeq
with the remaining thermodynamic quantities following from
Eq.~(\ref{usualrelations}).

Having thus verified the
low- and high-temperature scaling behaviors of our solutions,
we now seek to determine whether these solutions correctly match
the expected temperature dependence for {\it all}\/ temperatures.
Of course, to do this we must select a particular string model.

In general, for thermal duality invariant ground states,
$Z_{\rm string}(\tau,T)$ in Eq.~(\ref{Vdef}) takes the form
\beq
     Z_{\rm string}(\tau,T) = 
         Z_{\rm model}(\tau) Z_{\rm circ}(\tau,T)~.
\label{factor}
\eeq
The first factor $Z_{\rm model}$ is temperature-independent
and model-specific, while
$Z_{\rm circ}$ reflects the sum over Matsubara momentum
and winding states and takes the form
\beq
    Z_{\rm circ}(\tau,T) ~=~ \sqrt{ {\rm Im}\, \tau}\, 
    \sum_{m,n\in\IZ} \overline{q}^{p_R^2/2} q^{p_L^2/2}
\label{circ}
\eeq
where  $q\equiv \exp(2\pi i\tau)$ and
$p_{L,R}\equiv (ma \pm n/a)/\sqrt{2}$ with 
$a\equiv T/T_c$.
This is, of course, nothing but the 
standard circle partition function,
with the thermal duality symmetry 
taking the form ($a\leftrightarrow 1/a, m\leftrightarrow n$).
Although not every $Z_{\rm string}$ factors as in Eq.~(\ref{factor}), this form is the most 
general form preserving thermal duality invariance;  other forms will be discussed below.

For simplicity, let us begin in $D=2$ and focus
exclusively on the temperature dependence 
by taking $Z_{\rm model}=1$.
The associated cosmological constant 
for this ``model'' is thus
$\Lambda\equiv -\half \calM^2 \int_\calF [d^2\tau /({\rm Im}\,\tau)^2] Z_{\rm model}=
-\pi \calM^2/6$.
Remarkably, substituting Eq.~(\ref{circ}) into Eq.~(\ref{Vdef})
and integrating, we find {\it exact agreement}\/ with $\calV^{(\ell=2)}(T)$ 
for all values of $T$.
Thus, our closed-form $\ell=2$ solution exactly reproduces
the complete temperature dependence corresponding to this $D=2$ circle
compactification!
Indeed, the mathematical equivalence of these two expressions has been
known in other contexts for some 
time (see, {\it e.g.}\/, Refs.~\cite{identity,identity2,identity3}).

Mathematically, this is a rather surprising result:
the temperature dependence in Eq.~(\ref{circ}), which appears through $a\equiv T/T_c$
and takes the form of a sum of $\tau$-dependent exponentials,
is then integrated over the fundamental domain of the modular group.
Nevertheless, the net result of this integration is to produce the 
simple, closed-form result encapsulated within $\calV^{(\ell=2)}(T)$.
Indeed, we now see that this is the unique 
functional form which ensures
that {\it all}\/ thermodynamic quantities
are thermal duality covariant.

This agreement is remarkable for another reason as well.
Given its formulation, our 
our ``bootstrap'' derivation makes use of a powerful, 
all-orders non-perturbative duality symmetry.
By contrast, our ``bottom-up'' string calculation represents 
only a one-loop result.
The exact agreement between the two would therefore tend to suggest
that the one-loop result for this $D=2$ example is ``exact'', receiving
no further contributions at higher loops.

Let us now consider the case in higher dimensions $D>2$.
As might be expected, things are more complicated.
In general, for $D\not=2$, the analogue of taking the model-independent
simplification $Z_{\rm model}=1$ is to take 
$Z_{\rm model}=({\rm Im}\,\tau)^{1-D/2}$, since this overall prefactor
is model-independent and required by modular invariance in higher dimensions.
Inclusion of this $D$-dependent prefactor
has the net effect of reweighting the contributions
from each term in the $Z_{\rm circ}$ power series because they are now 
being integrating over the modular-group fundamental domain
with an altered measure.

Despite these changes, we find that
our solutions $\calV^{(\ell)}(T)$ with $\ell=D>2$
again successfully capture
the dominant temperature dependence of the resulting integrals.
Unlike the case with $D=\ell=2$, this agreement is only approximate
rather than exact.  Nevertheless, we find that this agreement
holds to within one or two percent over the entire temperature
range $0\leq T \leq \infty$.

Once again, this is a rather striking result, indicating that our
functional forms continue to capture the dominant
temperature dependence, even in higher dimensions.
Of course, for $D>2$, our closed-form
solutions and the above one-loop results do
not agree exactly.  However, given 
the precision with which the
one-loop results appear to match these functional forms,
it is natural to attribute the failure to obtain an exact
agreement for $D>2$ to the fact that our modular integrals are only
one-loop expressions.  
We thus are led to conjecture that
our functional forms $\calV^{(\ell)}(T)$ indeed represent the exact
solutions for the finite-temperature vacuum amplitudes,
even in higher dimensions,
and that these solutions emerge only when the
contributions from all orders in perturbation theory (and perhaps
even non-perturbative effects) are included.
Viewed from this perspective, it is perhaps all the more remarkable
that we found an exact agreement for $D=2$, suggesting that
the one-loop result is already exact in this special case.

Of course, these partition functions only represent simple 
circle compactifications, and do not correspond to actual
string models.  Nevertheless, the net effect
of inserting non-trivial partition functions $Z_{\rm model}(\tau)$ 
into these integrals is merely to change the  {\it subleading}\/ temperature
dependence in a model-dependent way.  Thus,
we conclude that the leading temperature dependence continues
to be captured by our universal solutions $\calV^{(\ell)}(T)$ to high
precision.  Moreover, if our conjecture is correct, then
we expect these subleading model-dependent contributions
to be washed out as higher-order contributions are included
in the perturbation sum.  These issues will be discussed 
in more detail in Ref.~\cite{I}.  

Of course, this conjecture would require not only a special
temperature dependence at each order in perturbation theory, 
but also a specific value of the string coupling which sets
the relative sizes of the perturbative contributions.
However, since thermal duality transformations necessarily involve
shifts of the string coupling beyond one-loop order, 
the two issues are closely related and it is possible 
that the string coupling is also fixed in such scenarios,
perhaps by other, non-perturbative effects.

It may seem remarkable that we can exploit thermal duality in order
to obtain explicit closed-form solutions for our thermodynamic quantities.
However, this is somewhat analogous to situations involving conformal   
invariance, where conformal symmetry can be used to completely fix
the form of certain string scattering amplitudes. 

These results also provide some justification for our original
motivation concerning the duality transformation properties of the entropy.
Even though the entropy generically transforms in a non-covariant manner,
and even though this does not give rise to a mathematical inconsistency,
we now see that string theory manages to find ground states which
nevertheless come very close to having entropies
which transform covariantly.   
As far as we are aware,
there is no other mathematical or conceptual reason why this should 
be the case.

Along these lines, we emphasize that the solutions $V^{(\ell)}(T)$ 
we have found must be interpreted correctly and with the proper caveats.  
As is well known, string theories exhibit 
exponentially growing densities of states which are thought to give
rise to a so-called Hagedorn transition at a temperature near
$T_c$.  Beyond this temperature, the string degrees of freedom 
are believed to change in a profound way, thereby eliminating the
relevance of the partition functions $Z_{\rm string}(T)$ 
as descriptions of the physics for temperatures beyond the Hagedorn temperature. 
We shall discuss the relation between these results and the Hagedorn
phenomenon below.
However, our goal in this paper has {\it not}\/ been to explore the
post-Hagedorn phase of string theory, but rather to explore the
consequences of thermal duality in the {\it pre}\/-Hagedorn phase.  
Indeed, despite the possible existence of a potential Hagedorn transition,
the pre-Hagedorn partition
functions $Z_{\rm string}(T)$ 
satisfy the thermal duality relation in Eq.~(\ref{dualityrelation}),
which we can view as an algebraic constraint on the functional form
$Z_{\rm string}(T)$ as a function of an arbitrary variable $T$.
Even though we expect $Z_{\rm string}(T)$ to describe physics only in the
pre-Hagedorn phase of theory, we can exploit this thermal duality relation 
in order to obtain closed-form expressions
for our thermodynamic quantities as functions of $T$, as we have done.
However, the resulting expressions, 
like $Z_{\rm string}$ itself, 
should be interpreted as physically relevant only in the 
pre-Hagedorn phase of the theory.

\begin{figure}[ht]
\centerline{
   \epsfxsize 3.0 truein \epsfbox {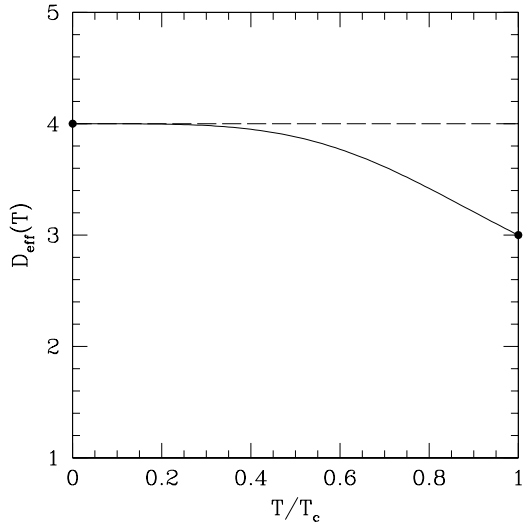}
    }
\caption{The effective dimensionality $D_{\rm eff}$ 
     of our four-dimensional thermodynamic solutions, 
    plotted as a function of $T$.
    These solutions behave ``holographically'', with
     the effective scaling dimensionality  
    falling exactly from $D_{\rm eff}=4$ to $D_{\rm eff}=3$.
    The dotted line indicates the behavior that would be expected
     within quantum field theory.}
\label{fig1}
\end{figure}

These closed-form solutions also have another intriguing property.
We have already remarked that the effective scaling behavior
of our solutions changes from $T^D$ at 
low temperatures to $T^2$ at high temperatures.
Given this, it is interesting to examine the effective dimensionality
(\ie, the effective scaling exponent)
of our solutions as a function of temperature.
In general, it is easiest to define this
effective dimensionality $D_{\rm eff}(T)$ by considering the entropy:
setting $S(T)\sim   T^{D_{\rm eff}-1}$, we obtain
\beq
    D_{\rm eff} ~\equiv~  1+ { d \ln S\over d\ln T} ~=~
               1 +  {T\over S} {dS\over dT} ~=~ 1 + {c_V\over S}~,
\label{Deff}
\eeq
where the last equality follows from the thermodynamic
identity $c_V= T dS/dT$.
Substituting our explicit solutions into Eq.~(\ref{Deff}), we find
the closed-form result 
$D_{\rm eff}(T) =  (2 T^D + D T_c^D)/(T^D + T_c^D)$.
Of course, as a consequence of thermal duality and the Hagedorn transition, 
the true ``high-temperature'' limit of string theory occurs 
not as $T\to \infty$, but as $T\to T_c$.
In this limit, we find $D_{\rm eff}\to \half (2 + D)$.
Thus, for $D=4$, 
we find that our solutions behave
exactly  ``holographically'' within the range $0\leq T\leq T_c$,
with the effective scaling dimensionality
falling from $D_{\rm eff}=4$ 
to $D_{\rm eff}=3$!~
A plot of $D_{\rm eff}$ as a function
of temperature is given in Fig.~\ref{fig1}.

%================== FIGURE ============================================
%================== END OF INSERTED FIGURE ============================

Of course, this is not true holography since our setup (based
on a flat-space computation) is incapable of yielding
the additional information concerning the geometry
associated with the surviving degrees of freedom that this
claim would require. 
Indeed, such a calculation would need to be performed
in a non-trivial background geometry in which both ``bulk''
and ``boundary'' are clearly identifiable. 
Nevertheless, we believe that this approach towards understanding
the relation between thermal duality and holography is
worthy of further investigation.

%=================================================================
\section{~Approach~\#2:\\  A duality covariant thermodynamics}
 
Despite the apparent successes of the bootstrap approach,
this method is less than satisfactory.
First, $c_V$ is not covariant unless $D\leq 2$.
But more importantly, 
because this approach towards restoring thermal duality
applies only for certain ground states,
it lacks the generality that should apply to a fundamental symmetry.
If thermal duality is to be considered an intrinsic property of
finite-temperature string theory (akin to T-duality), then the formulation
of the theory itself --- including its rules of calculation ---
should respect this symmetry regardless of the specific ground state.
This should be the case even if the particular string ground state is one
in which the thermal duality symmetry is spontaneously broken,
such as fermionic string theories (\eg, Type II or heterotic) 
where additional non-trivial
phases are required in 
$Z_{\rm string}$~\cite{Rohm,AlvOsoNPB304,McGuigan,AtickWitten,KounnasRostand},
thereby preventing $Z_{\rm string}$ from factorizing as in Eq.~(\ref{factor}).
After all, it is certainly acceptable if the
entropy or specific heat
fail to exhibit thermal duality because the ground state
fails to yield a
duality-symmetric
vacuum amplitude $\calV(T)$.   However, it is not acceptable if
this failure arises because the {\it definitions}\/ of the entropy or
specific heat in terms of $\calV(T)$ are themselves not duality covariant.

For this reason, we are motivated to develop a new,
alternative, fully covariant string thermodynamics
in which thermal duality is manifest.
Recall that the difficulty with the traditional rules of
thermodynamics in Eq.~(\ref{usualrelations})
is the temperature derivative:  as indicated in Eq.~(\ref{derivbad}),
the ordinary derivative $d/dT$ is not covariant with respect
to thermal duality transformations.  

Taking a clue from gauge
theory, let us therefore construct {\it a duality
covariant temperature derivative}\/.
In general, let us define $D_T\equiv d/dT + g(T)/T$ where
$g(T)$ is an unknown function.
Given the results in Eq.~(\ref{derivbad}), we immediately
see that if $f$ is a duality covariant function 
with weight $k$ and sign $\gamma$, then
$D_T f$ will be a duality covariant function
with weight $k-2$ and sign $-\gamma$ if and only if 
$g(T)+g(T_c^2/T)= -k$.

While many solutions for $g(T)$ may exist, let us take
thermal duality as our guide and demand that $g(T)$ is
itself a duality covariant function with weight $\alpha$ 
and sign $+1$.  We then find the unique solution 
$g(T)=  - k /[1 + (T_c/T)^\alpha]$, leading to the covariant
derivative
\beq
    D_T ~=~ {d\over dT} ~-~ {k\over T}\, 
       { T^{\alpha}\over T^{\alpha} + T_c^{\alpha} }~.
\label{Ddef}
\eeq
The analogy with gauge invariance is clear:  here the duality weight $k$ 
functions as the duality ``charge'' of the function being differentiated,
while the remaining factors are analogous to the connection. 
Together, the second term in Eq.~(\ref{Ddef}) functions as a 
``stringy'' correction term.
Although $\alpha$ is a free parameter, we shall restrict $\alpha>1$
in order to guarantee that this extra string contribution vanishes
in the low temperature $T\ll T_c$ limit.

Given this covariant derivative, we can now construct a manifestly
covariant thermodynamics:  our procedure is simply to replace
all temperature derivatives in Eq.~(\ref{usualrelations})
with the duality-covariant derivative in Eq.~(\ref{Ddef}).  We thus obtain
a manifestly covariant thermodynamics:
\beqn
          \tilde F = T \calV~,~~~~~~
         && \tilde U = - T^2 D_T \calV~,~~~\nonumber\\
         \tilde S = -D_T \tilde F~,~~~~~~ && \tilde c_V = D_T \tilde U~.
\label{newrelations}
\eeqn
The tildes emphasize that the new quantities
we are defining need not, {\it a priori}\/, be the same
as their traditional counterparts.
However, it is straightforward to verify that $\tilde F=F$ and $\tilde U=U$
for all temperatures, as expected.  The only modifications are in the definitions
of entropy and specific heat, which now obtain string-suppressed corrections
to their definitions:
\beq
      \tilde S = S +
              {2\, T^{\alpha-1} F \over T^\alpha+ T_c^\alpha}~,~~~~~
      \tilde c_V =  c_V -  {2 \,T^{\alpha-1} U\over T^{\alpha} + T_c^{\alpha} }~.
\label{newScv}
\eeq
Although these correction terms are suppressed for $T\ll T_c$,
they guarantee that $\tilde S$ and $\tilde c_V$ are thermal
duality covariant, as required.
Or, to phrase this result slightly differently, these corrections
ensure that the resulting thermodynamics is one in which all quantities
transform covariantly under the thermal duality map.
It is indeed remarkable that thermal duality covariance can
be explicitly restored to the standard rules of thermodynamics   
through the addition of correction terms which are all suppressed 
by powers of the string scale.

At a practical level, these extra terms pose no problems.
Since our string correction terms are all suppressed
for low temperatures, there is no experimental conflict 
with any branch of thermal physics, or with the 
standard theorems of thermodynamics, such as the second or third laws. 
Moreover, it is straightforward
to show that no obvious problematic issues of interpretation arise
at any temperature.
For example, the string-corrected specific heat $\tilde c_V$ 
always remains positive, and thus these 
corrections do not introduce thermal instabilities.

However, these modifications raise a number of important 
theoretical issues.
For example, by redefining entropy, 
we are clearly modifying the manner by
which one counts the number of statistical-mechanical
degrees of freedom, especially
at high temperatures near the string scale.
Can this be interpreted as indicative of some sort of breakdown
of the usual axioms of thermodynamics and statistical mechanics  
near the string scale?
After all, if new quantum-gravitational or string-induced effects 
ultimately distort the manner in which the system explores 
its energetically allowed states,
the string-corrected entropy may be precisely what accounts for
this phenomenon, providing a recipe for computing an ``effective''
number of degrees of freedom after all gravitational or string-induced
effects are included. 
These and other aspects of our string-corrected thermodynamics
will be discussed more fully in Ref.~\cite{II}.

It is, clearly, a big step to propose a modification to the laws
of thermodynamics, even if these corrections are essentially
unobservable at temperatures small compared to the string scale.
However, it may be argued that these corrections are, in some sense,
forced upon us in string theory.
In the case of geometric string compactifications,
we know that there are target-space dualities [such as $SL(2,\IZ)$ dualities]
which govern the resulting compactification physics, leading to
physical quantities (such as compactification superpotentials) which are 
$SL(2,\IZ)$ covariant.  Within such frameworks, it is well understood
that one must utilize $SL(2,\IZ)$-covariant
derivatives when differentiating with respect to compactification parameters.
If the correspondence between geometric 
compactification and finite-temperature effects is to have any validity
in string theory, then a similar thing must occur on the thermal side:
thermal quantities should be covariant with respect to thermal duality
transformations, and duality-covariant thermal derivatives should be employed
when taking temperature derivatives.  In this context, we note that this geometry/thermal
correspondence is more than a mathematical accident in string theory, for both
sets of symmetries ultimately have the same worldsheet origins.
For example, even if we had wished to formulate a finite-temperature version
of string theory without the introduction of thermal winding modes, such
a thing would not have been possible;  the resulting theory would 
necessarily break modular invariance, and consequently violate conformal
invariance at the one-loop level.
Thus the emergence of a thermal duality symmetry, and the resulting
need for a covariant string thermodynamics, is essentially unavoidable.

%=================================================================
\section{Beyond self-duality:  Spacetime fermions and the spontaneous 
breaking of thermal duality}

Thus far, we have
been focusing on situations which are in some sense
``self-dual'':  a single finite-temperature (bosonic) theory 
gives rise to vacuum amplitudes which are 
invariant under the algebraic replacement $T\to T_c^2/T$.  
However, the ideas developed in the
previous sections can also be directly applied to more realistic (fermionic)
string theories 
in which the thermal duality is spontaneously broken.  
It is important to consider these fermionic cases because these are the only 
known strings which are stable and tachyon-free.
In this section, we shall briefly sketch how the techniques in previous
sections can be extended to such cases.
More details can be found in Ref.~\cite{III}. 

Thus far, we have been assuming that all of our spacetime states are 
spacetime bosons, accruing integer-moded Matusbara excitations 
as in Eq.~(\ref{circ}).
However, as is well known, fermionic states must be moded 
with {\it half}\/-integer Matsubara
frequencies.  In string theory, this can be accomplished by
performing a $\IZ_2$ ``twist'' on the thermal Euclidean time/temperature
circle.  In general, this destroys the simple factorized form of
the thermal partition function as in Eq.~(\ref{factor}),
and spontaneously breaks the thermal (self-)duality.
We then find that the resulting thermal string model interpolates  
between two {\it different}\/ string models:  a given string model $M_1$
at temperature $T=0$ smoothly deforms to a {\it different}\/ string model $M_2$
at $T=\infty$.  Indeed, for such a string theory,
we find that     
\beq
         F(T) ~\sim~ \cases{  \Lambda_1 &  as $T\to 0$\cr
                              \Lambda_2 T^2/T_c^2 &  as $T\to \infty$ \cr}
\label{newFlimits}
\eeq
where $\Lambda_{1,2}$ are the one-loop cosmological constants of the
two different string models involved in the interpolation.   

It is important to interpret this interpolation correctly.
Of course, one interpretation is that model $M_1$ at temperature $T$ is dual
to a different model $M_2$ at temperature $T_c^2/T$;  in this sense,
such models are not ``self-dual'', but dual to each other.  However, 
we can also continue to view the entire interpolation as a description of 
the thermal behavior of the original model $M_1$ a function of $T$, 
as in Eq.~(\ref{newFlimits}).  It is in this
sense that we state that thermal duality is broken for Model~$M_1$. 

At first glance, the emergence of these non-trivial interpolations
would appear to invalidate the analysis of the previous 
sections.
However, somewhat surprisingly, we find that 
the techniques developed 
in the previous sections still continue to apply with only slight
modifications.  For example, it is 
straightforward to verify in such cases that the free 
energies $F(T)$ continue to transform as duality-covariant 
functions of weight $k=2$, but with respect 
to a {\it new, shifted}\/ critical temperature 
$T_\ast\equiv \sqrt{\Lambda_1/\Lambda_2} T_c$.   
Of course, while this is entirely consistent with the asymptotic behaviors
in Eq.~(\ref{newFlimits}), it was hardly required that such a shifted duality
symmetry hold for all temperatures.  

We can even go one step further.
Because the thermal duality is spontaneously broken in such scenarios,
we do not expect the traditional entropy to be transform covariantly at all,
and direct calculations in such theories verify this expectation.
However, using our string-corrected thermodynamics in Eqs.~(\ref{newrelations})
and (\ref{newScv}), we find that in many cases the {\it string-corrected entropies}\/ $\tilde S$
are actually invariant under the same shifted thermal duality    
with the same shifted temperature $T_\ast$!  
Given that the string corrections in Eq.~(\ref{newScv}) 
involve only the critical temperature $T_c$ in their definitions,
this is very surprising:  somehow the string corrections in 
Eq.~(\ref{newScv}), which are insensitive to the model-dependent parameter $T_\ast$,
manage to combine with the original uncorrected entropies
in order to render the full string-corrected entropy invariant with respect to the 
 {\it shifted}\/ thermal duality symmetry!

Clearly, this cannot occur unless the free energies $F(T)$ in such cases
have a very special algebraic form.
It is therefore possible to combine these results
in order to develop a new bootstrap solution
for $F(T)$ in such situations, obtaining a closed-form expression of the form
\beq
      F(T)= \Lambda_1  \left\lbrack 
                1 + \left( {T\over T_c}\right)^D\right\rbrack^{1/D}
            \left\lbrack
                1 + \left( {\Lambda_2\over \Lambda_1}\right)^{D}
\left( {T\over T_c}\right)^D\right\rbrack^{1/D}.
\eeq
As expected, this solution satisfies Eq.~(\ref{newFlimits}), and also
transforms with weight $k=2$ under $T\to T_\ast^2/T$ where $T_\ast$ is
defined as above.
This also reduces back to Eq.~(\ref{Ffullform}) 
in the self-dual case as $\Lambda_2\to \Lambda_1\equiv \Lambda$.
We stress, however, that this is only one of several possible bootstrap
formulations and solutions that can arise for such situations in which
thermal duality is spontaneously broken;  a more complete discussion can
be found in Ref.~\cite{III}.

Finally, we briefly situations in which the zero-temperature model
$M_1$ not only contains fermions, but is actually supersymmetric.
In such cases, $\Lambda_1=0$, and the above bootstrap results do
not apply.  However, even in such cases, it is possible to use
our string-corrected thermodynamics in order to uncover the hidden
thermal duality symmetries obeyed by $\tilde S(T)$, and to use these 
new symmetries in order to develop an appropriate bootstrap, in complete
analogy to what was done above.  Results along these lines will be
presented in Ref.~\cite{III}.  
However,
an important feature in such cases
is the fact that we no longer have $F(T)\to$~constant as
$T\to 0$.  Instead, since $\Lambda_1=0$, only the subleading behavior
$F(T)\sim \lambda_1 T^D/T_c^D$ as $T\to 0$ survives, where $\lambda_1$ is a subleading
coefficient.  Since such models continue to exhibit
the scaling behavior $F(T)\sim \Lambda_2 T^2/T_c^2$ as $T\to \infty$,
this implies that 
$F(T)$ in such cases can no longer transform with weight $k=2$
under any temperature inversion symmetry;  instead, we find that
we must have $k=2+D$ and $(T_\ast/T_c)^{D-2}\equiv \Lambda_2/\lambda_1$.
Thus, we see that 
in such cases, even the {\it duality weights}\/ of  
our thermodynamic quantities can be altered by spontaneous breaking
of the thermal duality symmetry.  
Nevertheless, our previous techniques will continue to apply.

%=================================================================
\section{Conclusions and open questions}

In this paper, we set out to address a very simple issue:
even though thermal duality is an apparent fundamental property of string
theory, emerging as a consequence of Lorentz invariance
and T-duality, the rules of classical thermodynamics
do not appear to respect this symmetry. 
In particular, they result in physically relevant 
thermodynamic quantities such
as entropy which fail to be duality covariant.

Demanding that string theory not give rise
to a duality non-covariant entropy,
we developed
a ``bootstrap'' approach towards obtaining the temperature
dependence of the effective potential for certain 
finite-temperature ground states. 
Remarkably, this yielded exact closed-form results
as well as others which we conjectured to be exact
when contributions from all orders in perturbation theory 
(and even non-perturbative effects) are included.
 
The existence of these closed-form solutions 
does not, however, address the fundamental problem that the
 {\it rules}\/ of thermodynamics are themselves non-covariant. 
We therefore proceeded to develop an alternative, manifestly covariant
thermodynamics which reduces to the traditional thermodynamics
at low temperatures, but which contains
significant modifications at higher temperatures near the string scale.

At a phenomenological level,
these speculations prompt many questions.
For example, it is important to further explore
how our closed-form results can be extended to theories in which thermal
duality is spontaneously broken (such as Type~II and heterotic
finite-temperature string ground states), as well as to
open strings and branes.
The results of such studies could have important
implications for recent brane-world scenarios, and are currently
underway~\cite{III}.
Likewise, it is interesting to consider the possible applications of
our results to early-universe cosmology, particularly regarding
the issues of phase transitions and entropy generation.

However, perhaps even more interesting are various theoretical 
questions that this approach raises.  
Can the string-corrected entropy continue to be interpreted 
as corresponding to disorder, or to heat transfer?
Can we develop an equivalent {\it microcanonical}\/ formulation 
of the modified thermodynamics we have developed here?
Indeed, is it even legitimate to tamper with the laws of 
thermodynamics near the string scale?
These issues will be discussed more fully in Ref.~\cite{II}.

Needless to say, there are many important aspects of string
thermodynamics which we have not touched upon.
These include the nature of the Hagedorn phase transition
as well as the Jeans instability and general issues 
concerning the interplay between gravity and thermodynamics. 
It will be interesting to explore the extent to which thermal
duality can shed light on these issues. 

There are two issues, in particular, which are directly
relevant to the viability of our proposals.
First, although we have focused on 
flat-space solutions of string theory, it is known that 
strings in AdS backgrounds can be reformulated as gauge field
theories~\cite{Maldacena}.  This implies that in such backgrounds,
string theories cannot give rise to the 
asymptotic behavior $F(T)\sim T^2$ 
that they exhibit in flat space, which would seem to invalidate
the existence of thermal duality as a fundamental symmetry of string
theory.
However, we have already noticed above that in certain situations,
spontaneous breakings of thermal duality can give rise to altered
asymptotic scaling behaviors, essentially deforming the 
duality weights of our thermodynamic quantities.
Thus, it is entirely possible that even in such AdS cases, 
the absence of thermal duality should more correctly be viewed 
as a spontaneous
breaking of thermal duality
induced by the altered
background geometry.
To the extent that such a reformulation is possible, 
we expect that some form of our
string-corrected thermodynamics should continue to be relevant.
However, the legitimacy of our approach clearly demands that
it be possible to formulate strings in such backgrounds as having
spontaneously broken thermal dualities.
These issues, which clearly include relations to black-hole thermodynamics
and D-brane counting, require further study.

Another important phenomenon in string thermodynamics
is the Hagedorn transition.  At first glance, it may seem
that our results (in particular, our closed-form solutions
in Sect.~III) are inconsistent with the existence of a Hagdorn 
transition since they provide smooth functions at all temperatures,
even as we approach the relevant Hagedorn temperatures from below
(where we trust our solutions to provide meaningful descriptions
of the relevant physics).
However, as has become increasingly clear in the recent string
literature, and as will be more fully discussed in Ref.~\cite{IV},
the usual Hagedorn transition is actually {\it absent}\/
for wide classes of string theories.  This occurs because modular
invariance provides an ultraviolet regulator which softens and
often eliminates the point-particle divergence that would have
arisen from an exponential rise in the degeneracy of string states.
Equivalently stated, the thermal ground-state winding mode  
which is normally assumed to become tachyonic at the Hagedorn 
temperature~\cite{AtickWitten}
is often GSO-projected out of the one-loop string partition function,
and does not give rise to divergences in the one-loop thermodynamic
quantities.
Thus, our results here are not in conflict with our normal
expectations as far as the Hagedorn transition is concerned. 
This issue will be discussed more fully in Ref.~\cite{IV}.

Throughout much of the past decade, 
progress in string theory has occurred through the study of
non-perturbative dualities.
Rather than dismiss these dualities as accidents of particular
string compactifications, we now regard such dualities as means
by which to gain insight into the truly ``stringy'' behavior
of physics at the most fundamental length scales.
In this paper, our goal has been to exploit thermal duality in much the same
way, to learn to something ``stringy'' about 
the possible nature of temperature, state counting, and thermodynamics 
near the string scale.
After all, if thermal effects can truly be associated with
spacetime compactification through the Matsubara/Kaluza-Klein correspondence,
then our expectations of an unusual ``quantum geometry'' near the string scale ---
one which does not distinguish between ``large'' and ``small'' ---
should simultaneously lead to expectations of an equally unusual thermodynamics
near the string scale which does not distinguish between ``hot'' and ``cold''
in the traditional sense.  Thermal duality should then serve as a tool
towards deducing the nature of these new effects.

%============================================================================= 
\section*{Acknowledgments}

This work was supported in part by the National Science Foundation
under Grants PHY-0071054 and PHY-0301998, and by a Research Innovation Award from 
Research Corporation. 
We wish to thank S.~Chaudhuri, E.~Dudas, C.~Kounnas, D.~Marolf, I.~Mocioiu,
R.~Myers, R.~Roiban, S.~Sethi, C.~Stafford, H.~Tye, and U.~van Kolck for discussions.

%=============================================================================

\medskip

%========================================================================
%========================================================================
%========================================================================


\begin{references}



\bibitem{Pol86}   J.~Polchinski,
     %``Evaluation Of The One Loop String Path Integral,''
     Commun.\ Math.\ Phys.\  {\bf 104} (1986) 37.
     %%CITATION = CMPHA,104,37;%%

\bibitem{McClainRoth}
      B.~McClain and B.~D.~B.~Roth,
      %``Modular Invariance For Interacting Bosonic Strings At Finite Temperature,''
      Commun.\ Math.\ Phys.\  {\bf 111} (1987) 539.
      %%CITATION = CMPHA,111,539;%%

\bibitem{OBrienTan}
      K.~H.~O'Brien and C.~I.~Tan,
      %``Modular Invariance Of Thermopartition Function And Global Phase Structure Of Heterotic String,''
      Phys.\ Rev.\ D {\bf 36} (1987) 1184.
      %%CITATION = PHRVA,D36,1184;%%

\bibitem{Polbook}
      For an introduction, see J. Polchinski, {\it String Theory, Vol.~I} 
      (Cambridge University Press, 1998), Chap.~9.


\bibitem{AlvOsoNPB304}
       E.~Alvarez and M.~A.~R.~Osorio,
       %``Cosmological Constant Versus Free Energy For Heterotic Strings,''
       Nucl.\ Phys.\ B {\bf 304} (1988) 327
       [Erratum-ibid.\ B {\bf 309} (1988) 220].
       %%CITATION = NUPHA,B304,327;%%


\bibitem{AtickWitten}
           J.~J.~Atick and E.~Witten,
           %``The Hagedorn Transition And The Number Of Degrees Of Freedom Of String Theory,''
           Nucl.\ Phys.\ B {\bf 310} (1988) 291.
           %%CITATION = NUPHA,B310,291;%%


\bibitem{AlvOsoPRD40}
        E.~Alvarez and M.~A.~R.~Osorio,
       %``Duality Is An Exact Symmetry Of String Perturbation Theory,''
       Phys.\ Rev.\ D {\bf 40} (1989) 1150.
       %%CITATION = PHRVA,D40,1150;%%


\bibitem{OsoIJMP}
         M.~A.~R.~Osorio,
         %``Quantum fields versus strings at finite temperature,''
         Int.\ J.\ Mod.\ Phys.\ A {\bf 7} (1992) 4275.
         %%CITATION = IMPAE,A7,4275;%%

\bibitem{shyam1}
       S.~Chaudhuri,
        %``Finite temperature bosonic closed strings: Thermal duality and the KT  transition,''
        Phys.\ Rev.\ D {\bf 65} (2002) 066008
        [arXiv:hep-th/0105110].
        %%CITATION = HEP-TH 0105110;%%


\bibitem{SakaiSenda}
       N.~Sakai and I.~Senda,
       %``Vacuum Energies Of String Compactified On Torus,''
       Prog.\ Theor.\ Phys.\  {\bf 75} (1986) 692
       [Erratum-ibid.\  {\bf 77} (1987) 773].
       %%CITATION = PTPKA,75,692;%%

\bibitem{Nairetal}
       V.~P.~Nair, A.~D.~Shapere, A.~Strominger and F.~Wilczek,
       %``Compactification Of The Twisted Heterotic String,''
       Nucl.\ Phys.\ B {\bf 287} (1987) 402.
       %%CITATION = NUPHA,B287,402;%%

\bibitem{Sathiapalan}
       B.~Sathiapalan,
      %``Duality In Statistical Mechanics And String Theory,''
      Phys.\ Rev.\ Lett.\  {\bf 58} (1987) 1597.
      %%CITATION = PRLTA,58,1597;%%


\bibitem{identity}
      M.~Bershadsky and I.~R.~Klebanov,
       %``Genus One Path Integral In Two-Dimensional Quantum Gravity,''
          Phys.\ Rev.\ Lett.\  {\bf 65} (1990) 3088.
       %%CITATION = PRLTA,65,3088;%%

\bibitem{identity2}
      N.~Sakai and Y.~Tanii,
       %``Compact Boson Coupled To Two-Dimensional Gravity,''
       Int.\ J.\ Mod.\ Phys.\ A {\bf 6} (1991) 2743.
       %%CITATION = IMPAE,A6,2743;%%


\bibitem{identity3}
       D.~J.~Gross and I.~R.~Klebanov,
      %``One-Dimensional String Theory On A Circle,''
      Nucl.\ Phys.\ B {\bf 344} (1990) 475.
      %%CITATION = NUPHA,B344,475;%%


\bibitem{I}  K.~R. Dienes and M. Lennek,
     {\it Adventures in Thermal Duality (I):  Extracting Closed-Form Solutions
        for Finite-Temperature Effective Potentials in String Theory},
          hep-th/0312216 (Phys.\ Rev.\ D, in press).


\bibitem{Rohm}
      R.~Rohm,
      %``Spontaneous Supersymmetry Breaking In Supersymmetric String Theories,''
      Nucl.\ Phys.\ B {\bf 237} (1984) 553.
      %%CITATION = NUPHA,B237,553;%%


\bibitem{McGuigan}
     M.~McGuigan,
     %``Finite Temperature String Theory And Twisted Tori,''
     Phys.\ Rev.\ D {\bf 38} (1988) 552.
     %%CITATION = PHRVA,D38,552;%%


\bibitem{KounnasRostand}
      C.~Kounnas and B.~Rostand,
      %``Coordinate Dependent Compactifications And Discrete Symmetries,''
      Nucl.\ Phys.\ B {\bf 341} (1990) 641.
      %%CITATION = NUPHA,B341,641;%%

\bibitem{II} K.~R. Dienes and M. Lennek,
   {\it Adventures in Thermal Duality (II):  
          Towards a Duality-Covariant String Thermodynamics},
         hep-th/0312217 (Phys.\ Rev.\ D, in press).

\bibitem{III} K.~R. Dienes and M. Lennek,
   {\it Adventures in Thermal Duality (III):  
          Beyond Spontaneous Breaking:  The Re-Emergence
           of Thermal Duality in Fermionic String Theories},
        to appear.


\bibitem{Maldacena}
      J.~M.~Maldacena,
      %``The large N limit of superconformal field theories and supergravity,''
      Adv.\ Theor.\ Math.\ Phys.\  {\bf 2} (1998) 231
      [Int.\ J.\ Theor.\ Phys.\  {\bf 38} (1999) 1113]
      [arXiv:hep-th/9711200].
      %%CITATION = HEP-TH 9711200;%%


\bibitem{IV}
        K.~R. Dienes and M. Lennek,
       {\it Re-Identifying the Hagedorn Transition}, to appear.
      


\end{references}
\end{document}